\documentclass[conference]{IEEEtran}
\IEEEoverridecommandlockouts

\usepackage{amsmath,amsfonts}
\usepackage{algorithm}
\usepackage{xkeyval}
\usepackage{graphicx}
\usepackage{textcomp}
\usepackage{xcolor}
\usepackage{url}
\usepackage{multirow}
\usepackage{tabularx}
\usepackage{comment}
\usepackage{booktabs}
\usepackage{soul}
\usepackage[noend]{algpseudocode}
\usepackage{xspace}
\usepackage{enumitem}
\usepackage{fancyvrb}
\usepackage[english]{babel}
\usepackage{blindtext}
\usepackage{csquotes}
\usepackage{caption}
\usepackage{svg}
\usepackage{float}
\usepackage[draft]{hyperref}

\usepackage[top=0.75in, bottom=1.1in, left=0.625in, right=0.625in]{geometry}

\def\BibTeX{{\rm B\kern-.05em{\sc i\kern-.025em b}\kern-.08em T\kern-.1667em\lower.7ex\hbox{E}\kern-.125emX}}

\newcommand{\sysname}{MobiLLM\xspace}

\newenvironment{packeditemize}{
\begin{list}{$\bullet$}{
\setlength{\itemsep}{1.5pt}
\setlength{\labelwidth}{8pt}
\setlength{\leftmargin}{10pt}
\setlength{\labelsep}{3pt}
\setlength{\listparindent}{\parindent}
\setlength{\parsep}{1.5pt}
\setlength{\parskip}{1.5pt}
\setlength{\topsep}{1.5pt}}}{\end{list}}

\let\OLDthebibliography\thebibliography
\renewcommand\thebibliography[1]{
  \OLDthebibliography{#1}
  \setlength{\parskip}{4pt}
  \setlength{\itemsep}{0pt plus 0.3ex}
}

\begin{document}


\title{\sysname: An Agentic AI Framework for Closed-Loop Threat Mitigation in 6G Open RANs}

\author{
Haohuang Wen$^{*23}$\thanks{The first two authors contributed equally to this work}, Prakhar Sharma$^{*12}$, 
Vinod Yegneswaran$^{12}$, Ashish Gehani$^{12}$, Phillip Porras$^{12}$, Zhiqiang Lin$^{23}$ \vspace{0.3em} \\

{\normalsize $^{1}$SRI}, 
{\normalsize $^{2}$SE-RAN.ai}, 
{\normalsize $^{3}$The Ohio State University} \\

{\normalsize \{prakhar, haohuang, phil\}@se-ran.ai},
{\normalsize \{vinod.yegneswaran, ashish.gehani\}@sri.com}, 
{\normalsize \{lin.3021\}@osu.edu}
}

\maketitle

\begin{abstract}
The evolution toward 6G networks is being accelerated by the Open Radio Access Network (O-RAN) paradigm—an open, interoperable architecture that enables intelligent, modular applications across public telecom and private enterprise domains. While this openness creates unprecedented opportunities for innovation, it also expands the attack surface, demanding resilient, low-cost, and autonomous security solutions. Legacy defenses remain largely reactive, labor-intensive, and inadequate for the scale and complexity of next-generation systems. Current O-RAN applications focus mainly on network optimization or passive threat detection, with limited capability for closed-loop, automated response.

To address this critical gap, we present an agentic AI framework for fully automated, end-to-end threat mitigation in 6G O-RAN environments. MobiLLM orchestrates security workflows through a modular multi-agent system powered by Large Language Models (LLMs).
The framework features a \textbf{Threat Analysis Agent} for real-time data triage, a \textbf{Threat Classification Agent} that uses Retrieval-Augmented Generation (RAG) to map anomalies to specific countermeasures, and a \textbf{Threat Response Agent} that safely operationalizes mitigation actions via O-RAN control interfaces. Grounded in trusted knowledge bases such as the MITRE FiGHT framework and 3GPP specifications, and equipped with robust safety guardrails, \sysname provides a blueprint for trustworthy AI-driven network security. Initial evaluations demonstrate that \sysname can effectively identify and orchestrate complex mitigation strategies, significantly reducing response latency and showcasing the feasibility of autonomous security operations in 6G.

\end{abstract}

\section{Introduction}\label{Introduction} 

The rise of sixth-generation (6G) wireless networks is tightly coupled with the Open Radio Access Network (O-RAN) paradigm\cite{o-ran}--a shift toward openness that is transforming the telecom network ecosystem. By enabling third-party developers to deliver modular xApps and rApps that interoperate across vendors, O-RAN unlocks rapid innovation and paves the way for 6G adoption in mission-critical domains such as remote healthcare, smart manufacturing, and autonomous transportation. Yet, this open, multi-vendor environment also introduces major challenges. The cost to manage such a complex network is expected to be extremely high, due to its sophisticated protocols and the need for deep operational expertise. Meanwhile, 6G networks will likely inherit known vulnerabilities from prior standards—including over-the-air attacks from malicious devices, rogue base stations, and man-in-the-middle threats—while also experiencing new run-time faults like registration failures, software misconfigurations, and other unexpected behaviors~\cite{kim2019touching,hussain20195greasoner,lee2019your,bitsikas2022you,kotuliak2022ltrack,erni2022adaptover}. If not addressed properly, these persistent threats and operational instabilities will weaken the trust required to deploy 6G in security-critical domains.

The current research landscape reveals two parallel streams of work in O-RAN: passive security detection and non-security-focused automation. On one hand, significant research has concentrated on threat detection. This includes various xApps designed to identify network anomalies~\cite{sun2024spotlight} and malicious activities such as the presence of rogue UEs and base stations~\cite{scalingi2024det,5G-Spector:NDSS24,huang2023developing,wen20246g}. A critical limitation of these solutions, however, is their passive nature; they are designed to report threats but lack the capability to automatically respond or mitigate them. On the other hand, numerous closed-loop xApps and rApps have been developed for network automation and performance optimization tasks like network slicing and resource management~\cite{polese2022colo,lacava2023programmable,d2022orchestran,yeh2023deep}. These systems, while demonstrating the potential of automation, were not created to address cybersecurity challenges. This leaves a crucial gap for a framework that can unify intelligent detection with proactive, closed-loop security response.

More recently, Large Language Models (LLMs) have emerged as a promising tool for O-RAN network management, such as network slicing and resource management ~\cite{oranguide, llmxapp}. While these studies demonstrate the applicability of LLMs for network orchestration~\cite{endToEndOran,hric}, they exhibit several limitations that prevent them from filling the aforementioned security gap.
First, none of these LLM-based frameworks address the crucial aspect of closed-loop security management. An effective security system requires more than simple anomaly detection; it needs a sophisticated analysis engine grounded in the latest threat intelligence and an intelligent response engine capable of enacting changes to the O-RAN network configuration. Second, most of these works rely on basic prompt engineering with general-purpose LLMs, a method prone to hallucinations and irrelevant output when dealing with the highly technical and standardized nature of cellular networks. This highlights the need for robust knowledge grounding to ensure accurate, task-specific outputs. Consequently, the challenge of creating a framework that can autonomously detect, analyze, and respond remains unsolved. 

In this paper, we present \sysname to address these gaps by introducing a knowledge-grounded, multi-agent framework dedicated to closed-loop security response in O-RAN environments. However, pre-trained models, despite their impressive general capabilities, lack the deep, domain-specific knowledge contained within the extensive and highly technical specifications that govern cellular networks. Hence entrusting security-related tasks to an LLM is inherently risky. Unlike human experts, LLMs can hallucinate—generating plausible but false information, which could lead to incorrect and potentially disruptive mitigation actions in a live network. Therefore, the design of \sysname focuses on implementing robust guardrails that ensure its trustworthiness and operational safety.

To achieve this, we introduce two key innovations: first, to leverage the LLM as a powerful natural language parser that maps real-time threat descriptions to the structured MITRE FiGHT framework~\cite{mitre_fight}, and second; employing the ReAct pattern ~\cite{react} to translate the high-level mitigation strategy from FiGHT into a sequence of calls to a small, predefined set of safe, human-written APIs. This ensures that all interactions with the network are constrained to vetted, predictable operations, making the system's behavior both effective and verifiable.

\noindent{\bf Contributions. } Our main contributions include:
\begin{itemize}
    \item Development of \sysname, the first multi-tiered agentic framework for cellular threat analysis, planning, and response~\footnote{MobiLLM has been released at \url{https://github.com/5GSEC/MobiLLM}}.
    \item Evaluation of \sysname with a suite of five real-world 5G threat scenarios. We release associated RRC protocol traces and initial incident reports for each threat.
    \item Discussion of guardrails to prevent misaligned or hallucinated LLM outputs from affecting changes in the network.
\end{itemize}
\section{System Design and Implementation}\label{system_design}

\begin{figure*}[t]
    \centering
    \includegraphics[width=0.88\textwidth]{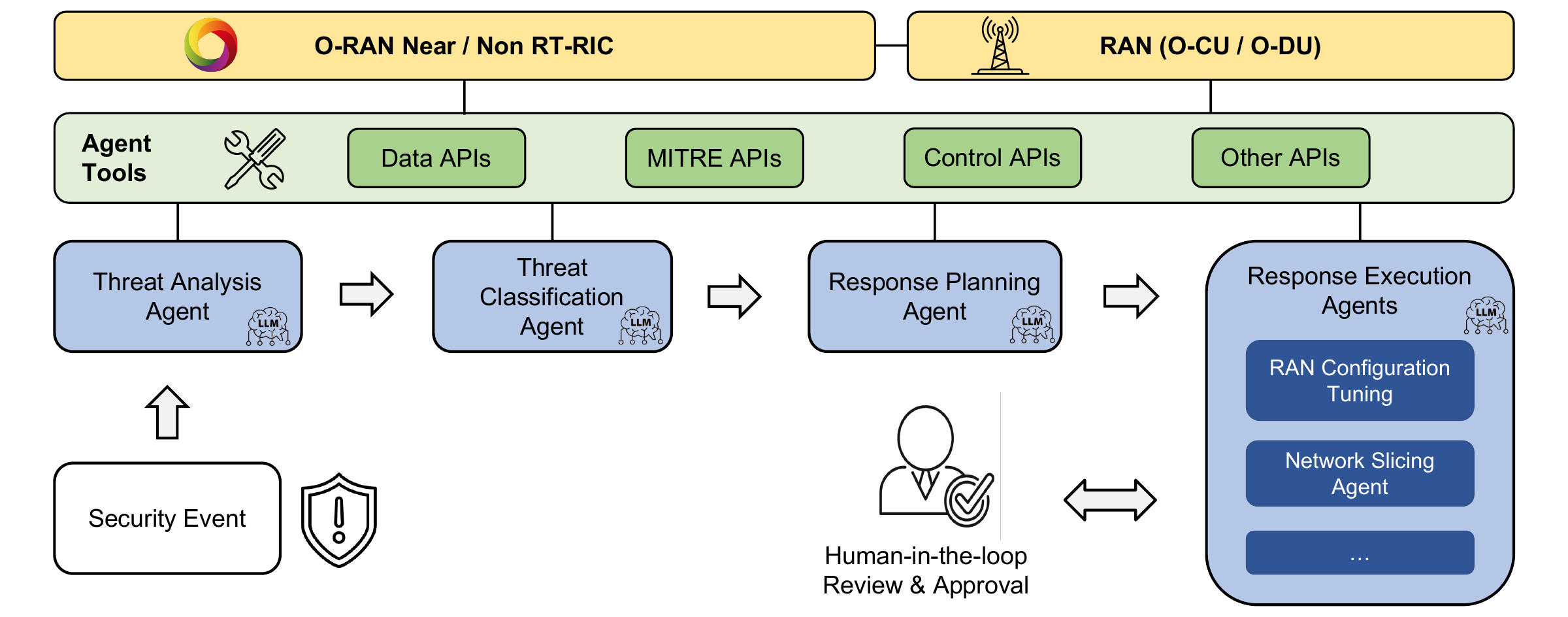}    
    \caption{\small Overview of \sysname's agentic architecture. The config tuning agent can be invoked autonomously, but the effects are only applied after human approval is gathered.} 
    \vspace{-0.1in}
\end{figure*}

We present the overall design of \sysname, a modular agentic AI framework created to manage the complex security needs of 6G networks. Applying existing LLM-based agentic frameworks directly to the high-stakes, dynamic 6G environment is not feasible, as mobile networks require highly specialized knowledge, an area where generic pre-trained LLMs are deficient. For instance, a recent GSMA report highlights that even state-of-the-art models like GPT-4 struggle to answer highly technical questions related to 3GPP specifications and telecommunication protocols~\cite{gsma_benchmark}.

To overcome this critical knowledge gap, \sysname is built as a multi-agent architecture that augments LLM capabilities with explicit domain awareness. The framework decomposes the complex security pipeline, i.e., from threat analysis to closed-loop mitigation, into three distinct stages. Each stage is handled by a specialized LLM-based agent that operates in sequence, creating a streamlined workflow. By assigning each agent a specific subtask with a precisely crafted prompt, our design maximizes accuracy and efficiency, ensuring that the system's reasoning and actions are grounded in the specific context of 6G network operations. 

\vspace{0.1in}
\noindent{\bf Deployment Scenarios.}
A potential deployment scenario for \sysname is within the O-RAN's Service Management and Orchestration (SMO) infrastructure~\cite{polese2022understanding}, where it can be instantiated as a modular rApp while not compromising the overall network latency in a non-critical network path. Within this architecture, \sysname is positioned to work synergistically with a wide range of threat and anomaly detection xApps and rApps deployed in the O-RAN control plane~\cite{scalingi2024det,5G-Spector:NDSS24,huang2023developing,sun2024spotlight}, and perform event analysis, classification, as well as closed-loop mitigation. The major interactions are conducted through the standard O-RAN interfaces such as O1 and A1, to seamlessly work with other network components including xApps, rApps, O-CU, and O-DU. The following describes the detailed design of \sysname.

\subsection{Threat Analysis Agent}

The Threat Analysis Agent serves as the initial triage system. Its primary input is not just a simple prompt, but a detected threat or abnormal event, which could originate from network monitoring tools, an O-RAN xApp/rApp, or a natural language query from a human operator. Upon receiving an alert, the agent's first moves is beyond the surface-level description by actively inspecting the event and correlating it with related network data, querying sources such as performance metrics, RAN-level data, and logs. This process allows the agent to analyze the context, evaluate the potential risk, and distinguish a genuine security threat from a benign network fault or false positive. If a credible threat is identified, the agent generates a structured, machine-readable threat report containing the event summary, affected components, and an initial risk assessment. If no threat is detected, it logs the analysis and closes the event, ensuring that downstream agents do not expend unnecessary computational resources.

\subsection{Threat Classification Agent}

The main goal of the Threat Classification Agent is to solve a critical problem with using LLMs for security: their lack of specialized knowledge and their tendency to hallucinate. Our novel approach doesn't trust the LLM to invent solutions. Instead, we use it as an intelligent {\it translator} to ground its reasoning in a reliable, human-curated knowledge base.
The key innovation is leveraging the MITRE FiGHT framework~\cite{mitre_fight} as this trusted source. This agent bridges the gap between a potentially ambiguous threat alert and a concrete, verifiable action plan. By forcing the agent to map every threat to a standard FiGHT technique, we ensure its classifications are accurate, consistent, and based on industry best practices. This transforms the LLM from an unreliable narrator into a dependable reasoning engine. The agent follows a systematic, two-phase process to achieve this:

\begin{packeditemize}
    \item {\bf Phase-I: Offline Knowledge Base Preparation}. First, we crawl the entire public MITRE FiGHT framework to gather all attack techniques and their detailed descriptions. This text data is then processed by a text embedding model, which converts the description of each technique into a numerical vector. Finally, these vectors are stored and indexed in a specialized vector database, creating a searchable library of 6G threats.

    \item {\bf Phase-II: Run-time Threat Classification}. The agent receives a structured threat report from the Threat Analysis Agent. It uses a Retrieval-Augmented Generation (RAG) process. The natural language description from the report is used as a query to search the vector database. The search retrieves the top matching MITRE FiGHT technique(s) based on semantic similarity. The agent then outputs a formal classification using the FiGHT technique ID and, most importantly, extracts the mitigation guidance associated with that technique, providing a clear plan for the next agent in the chain. 
    
\end{packeditemize}

\subsection{Response Planning and Execution Agent}

The final agent in the \sysname pipeline is responsible for translating a mitigation plan into safe, real-world action. The central motivation behind its design is to solve the most critical challenge in autonomous security: how to grant an LLM control over a live network without exposing the system to the risks of hallucination or unsafe command execution.
Our novel contribution is a two-layer architecture that separates high-level planning from low-level execution. We use the LLM as a sophisticated planner that reasons about what to do, but we strictly limit how it does it. The LLM's creative capabilities are used to dynamically construct an action plan by mapping the mitigation guidance from MITRE FiGHT~\cite{mitre_fight} to a set of available network control APIs. However, the final execution is delegated to specialized {\it Action Agents}, which operate within rigid, pre-defined workflows. An example of such includes a {\it Config Tuning Agent} that specifies a workflow to update the RAN configuration and perform a reboot to let it take effect. This design provides the flexibility of LLM-based planning while enforcing the safety of deterministic, human-vetted code, ensuring that every action taken is both intelligent and trustworthy. The agent executes its task through the following step-by-step process:

\begin{packeditemize}
    \item {\bf Action Plan Generation.} The agent receives the mitigation goal from the previous stage. It then analyzes its library of available, human-written safe APIs and generates a step-by-step plan to achieve the goal. If a viable plan cannot be constructed using only the provided tools, it proceeds directly to Step 3.
    
    \item {\bf Execution via Safe Workflows.} If a valid plan is created, it is dispatched to the appropriate specialized Action Agents. Each of these agents follows its own pre-defined, hard-coded workflow to execute its part of the plan. This critical safety guardrail ensures the LLM can only provide parameters while the rigid script controls the execution logic (the how), preventing any unsafe or unintended operations.
    
    \item {\bf Revert on Failure.} If no valid plan could be created in the first step, the agent does not attempt to improvise. Instead, it flags the issue and presents the original mitigation guidance as a high-level recommendation to a human operator, ensuring that a human is always in the loop for complex or unsupported threats.
\end{packeditemize}

\subsection{Implementation Details}

Following the aforementioned architecture of \sysname, we now describe the implementation details.

\vspace{0.1in}

\noindent \textbf{Knowledge Grounding. } We base the design of \sysname's agentic architecture on two recent LLM advancements: ReAct agents \cite{react}, and retrieval-augmented generation (RAG). The former is a few-shot prompting technique that has demonstrated good results in reasoning and action-based (tool calling) tasks. The latter is a technique to provide inference-time context to the LLM, and helps grounding the LLM response in existing verified knowledge bases and curb issues like hallucinations, meandering output and ill-formed responses. We employ numerous sources as RAG knowledge bases viz. MITRE FiGHT~\cite{mitre_fight}, 3GPP standard specifications, O-RAN documents and curated academic research papers documenting relevant 5G/LTE attacks and their mitigations. The main ones that interface with respective agents are described below:
\begin{packeditemize}
    \item \textbf{Network Data APIs: } These enable the threat analysis agent to gain an in-depth insight about the current network status at various levels of granularity. (e.g., fetching network traffic, logs, and detected events)
    \item \textbf{MITRE APIs: } The MITRE FIGHT framework \cite{mitre_fight} documents known 5G / 6G tactics, techniques and mitigations. This knowledge base of adversary behaviors is recorded as a database of popular techniques, their operational modalities, target attack surfaces and proposed mitigations. For any identified threat, this database is one of the most direct resources for crafting a response. These APIs enable the threat classification agent to search the MITRE FIGHT database for relevant tactics and techniques (and associated procedures and mitigations) given a certain security incident and threat analysis.
    \item \textbf{Control APIs:} These functions enable the RAN configuration tuning agent to perform an action in a human-supervised control loop. An agent can fetch the active configuration of a CU, suggest modifications, and, upon receiving human approval, apply the updated configuration and reboot the RAN to enact the changes. The scope of controllable parameters can be readily expanded by incorporating additional parameters from the O-RAN E2SM-RC (RAN Control) standard~\cite{polese2022understanding}.
\end{packeditemize}


\noindent \textbf{Agent Tools and Prompts. } The threat analysis, threat classification, security planning and action agent interface with the knowledge bases, RAN control system, configuration tuning subroutines and system reboot functions to dynamically analyze and craft a response to an ongoing threat. The agents are free to call tools in any fashion (single-turn, multi-turn, multi-step, etc.). We use React-style few-shot prompts that encourage the model to think, reason, and act in an iterative loop. Max iterations for that loop is five.

\noindent \textbf{Guardrails: } To address LLM issues viz. hallucinations, irrelevant outputs, and misinterpretation of contextually similar inputs, we incorporate multi-layered guardrails designed to ensure robustness, correctness, and operational safety:
\begin{packeditemize}
    \item {\bf Prompt-Level Guardrails}: Carefully crafted few-shot prompts are used to constrain the LLM’s output to a predefined structure.
    \item {\bf Rule-Based Guardrails}: Deterministic output checkers and JSON sanitizers are applied to filter out hallucinated or malformed outputs and enforce schema compliance.
    \item {\bf Human-in-the-Loop Validation}: Given the high-stakes nature of integration with systems like 6G networks, human feedback is employed to review and validate LLM-generated mitigation scripts. This step ensures that unintended or unsafe logic is not propagated into production environments.
\end{packeditemize}
\section{Evaluation}\label{evaluation}

\begin{figure}[!t]
    \centering
     
    \hspace*{-0.2in}
    \includegraphics[width=0.48\textwidth]{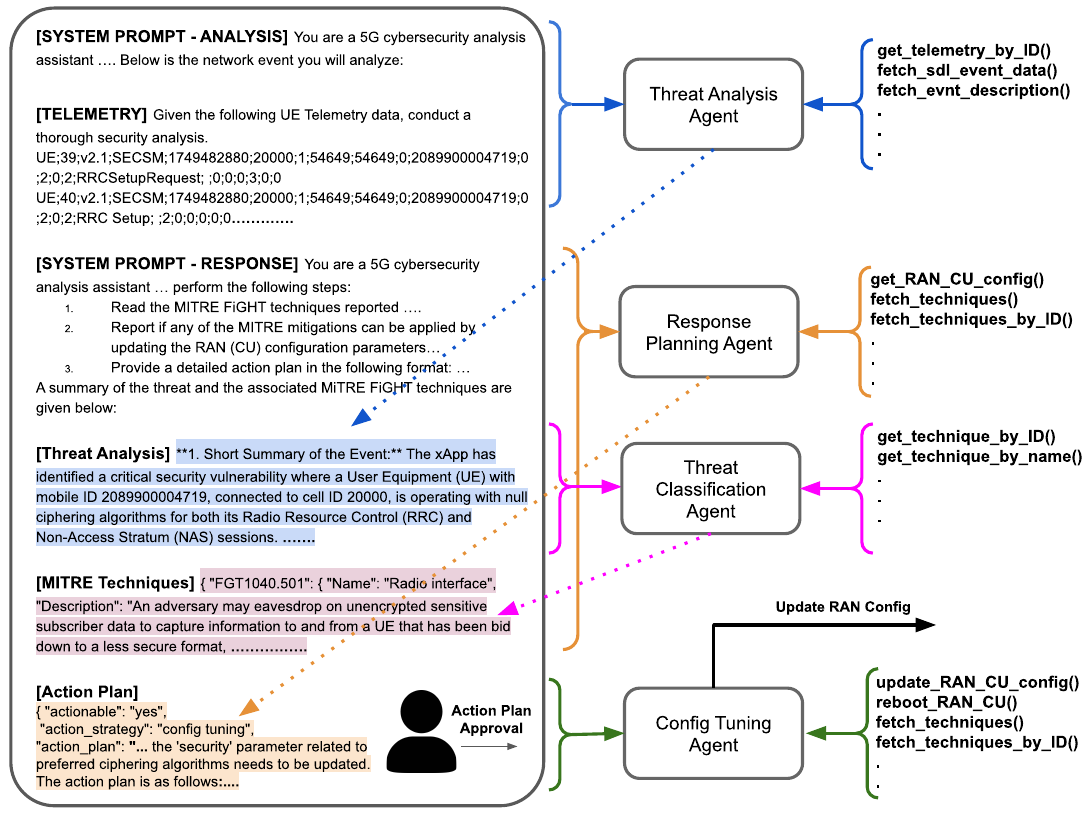}
    \label{fig:prompts} 
    \caption{\small This figure illustrates the interaction flow of \sysname, where specialized agents sequentially process incident telemetry to classify a threat and generate a response plan. The resulting configuration change is then presented for mandatory human approval before being executed by the appropriate tool.}
    \label{fig:example}
\end{figure}

Since \sysname represents the first agentic architecture capable of analyzing, classifying, and autonomously responding to threats within 6G O-RAN networks, defining a robust evaluation methodology posed a significant challenge. In the absence of a standardized dataset or established evaluation protocol for this domain, we curated a focused set of five representative attacks that 6G networks inherit from prior standards (e.g., LTE and 5G NR): Blind Denial-of-Service (DoS), Base Transceiver Station (BTS) resource depletion, Downlink IMSI exposure, Uplink IMSI exposure, and Null cipher exploitation~\cite{kim2019touching,erni2022adaptover,hussain20195greasoner}, which were reproduced end-to-end within a testbed using open-sourced software including OpenAirInterface and OpenRAN Software Community RIC.

\subsection{Evaluation Setup}
We construct a 5-case evaluation suite representative of the aforementioned 6G attacks. For each threat we record (i) A threat scenario describing the threat and expected observables; (ii) Curated telemetry (RRC and NAS traffic telemetry) corresponding to the attacks; and (iii) The MITRE FiGHT \cite{mitre_fight} tactics classification associated with each threat and their corresponding mitigations. We use the scenario description and the curated telemetry as context with the initial prompt to the threat analysis agent. To evaluate the threat classification performance, we ask a few cellular network experts to generate ground truth results for all the targeted attacks.

We evaluate \sysname using the following metrics: (1) MITRE FIGHT technique retrieval accuracies (2) Agent tool invocation validity: whether agent tools were recommended in the correct scenario (3) Overall remediation validity, which are validated by cellular network experts. We additionally report end-to-end latency per framework invocation.

\begin{table}[!h]
\centering
\scalebox{1.00}{
\begin{tabular}{lccc}
\hline
\textbf{Threat} & \textbf{Top-3} & \textbf{Top-1} & \textbf{CCR} \\
\hline
BTS-Attack-1            & 1 & 0.6 & 0.8\\
BTS-Attack-2            & 1 & 1 & 0.4\\
BTS-Attack-3            & 1 & 0.6 & 0.8\\
Blind-DoS-1             & 1 & 1 & 1\\
Blind-DoS-2             & 1 & 1 & 1\\
Blind-DoS-3             & 1 & 1 & 0.6\\
Downlink-DoS            & 1 & 0.4 & 0.4 \\
Downlink-IMSI           & 0.8 & 0 & 0\\
Null-Cipher-Integrity   & 0.8 & 1 & 1\\
Uplink-IMSI             & 0.8 & 0.4 & 0.4 \\
\hline
Mean                 & 0.94      & 0.72 & 0.64 \\
\hline
\end{tabular}
}
\caption{\small Per-sample MITRE FIGHT Technique retrieval performance (Top-3 and Top-1 accuracy), and Per-threat tool correct calling ratio (CCR) across five \sysname runs}
\label{tab:per_sample}
\end{table}



\subsection{Quantitative Results}

Our per-attack evaluation, summarized in \autoref{tab:per_sample}, reveals that \sysname achieves high accuracy in threat classification. Specifically, the correct MITRE FiGHT mitigation technique was present in the Top-3 retrieved results 94\% of the time. However, this success in classification does not fully translate to the action phase. As shown in \autoref{tab:per_sample}, the accuracy for correctly invoking the final mitigation action is considerably lower, at an average of 64\%. To investigate the root causes of this performance gap, we analyze two representative case studies in the following section.

\subsection{Overall Remediation Validity}
To illustrate \sysname's remediation capabilities and limitations, we analyze two representative case studies: one successful mitigation and one failure.

\noindent \textbf{Null Cipher-Integrity Attack.} 
A Null Cipher-Integrity attack occurs when a misconfigured base station forces a UE to disable confidentiality (EEA0) or integrity protection (EIA0) during its security setup, exposing user data. In our evaluation, \sysname executed a successful, human-supervised remediation workflow, as in \autoref{fig:example}. First, the Threat Classification Agent correctly mapped the anomaly to the corresponding MITRE FiGHT technique, `FGT1600.501' (Disabling Encryption Over Radio Interface). Based on this classification, the Response Planning Agent used the {\it get\_ue\_description tool} to gather context and correctly drafted a plan to remove the insecure `nea0' and `nia0' algorithms from the CU's configuration. Before execution, the agentic workflow paused for mandatory human approval. Once granted, the Config Tuning Agent successfully applied the changes using the {get\_ran\_cu\_config} and {\it update\_ran\_cu\_config} tools, resolving the underlying vulnerability.

\noindent \textbf{Downlink IMSI Extraction Attack.} A Downlink IMSI Extraction attack involves a man-in-the-middle attacker who intercepts and modifies a pre-authentication RRC message, forcing the UE to reveal its permanent identity (IMSI) in plain text before security is established~\cite{erni2022adaptover}. While \sysname again succeeded in classifying the attack with the correct MITRE FiGHT techniques, it failed during response planning. The agent mistakenly concluded that the threat could be mitigated by enforcing integrity protection on the connection. This plan was fundamentally flawed because the attack exploits a timing vulnerability that occurs before the authentication stage where such protections are activated. The root cause of this error lies in the limitations of the general-purpose LLM (Gemini 2.5 Flash) used in our evaluation, which lacks deep, specialized knowledge of the 3GPP state machine and specifications. This failure highlights the necessity of fine-tuning LLMs with domain-specific awareness to handle sophisticated cellular threats, a direction we explore in the discussion section.



\section{Discussion and Future Work}\label{sec:discussion}

\subsection{General purpose vs. specialized agents} \sysname is a promising direction towards fully automated threat incidence analysis and response. We perform all our experiments with Gemini-2.5-Flash as the core deduction engine for both analysis generation and function calling, and the distinction between agents is made through what tools they have at their disposal (geared towards a specific task). While knowledge bases and tools help ground the LLM response, further fine tuning is imperative for state-of-the-art performance for the following reasons:
\begin{packeditemize}
    \item API-based off-the-shelf models are ephemeral and costly. Training bespoke LoRA \cite{lora} adapters enables light-weight SLMs \cite{slm} to achieve better performance in agentic tasks. We plan to bootstrap the current capabilities of our Agentic architecture to create a supervised fine-tuning dataset; to train small but specialized language models specifically for crafting a threat response in the context of O-RAN 5G. 
    \item Currently, we supply network facts like CU configurations and parameters at inference time, which makes it slow and costly. Through supervised fine tuning, this information can be baked into the model parameters. We also plan to expand the use case to more than 5 attacks that we currently support, with tools designed to address a wide range of threats.
    \item Existing work has been successful in using telecom-specific knowledge to fine-tune an LLM to enhance domain awareness and understanding by fine-tuning on the vast corpus of 3GPP and O-RAN specifications~\cite{zou2025telecomgpt}. In the security context, we can create a true network security expert with deeper protocol understanding and more accurate tool-use capabilities than any general-purpose LLM. 
\end{packeditemize}

\subsection{Accessible and robust O-RAN security } We plan to integrate \sysname into real-world 6G deployments. As initiatives like the U.S. government's Open Centralized Unit Distributed Unit (OCUDU) project accelerate the creation of open 6G platforms, \sysname can be embedded as a core security layer to not only enhance network resilience and trust from the outset, but also cut down network operator training cost. We plan to develop more advanced security guardrails to protect \sysname from adversarial attacks like backdoor attacks \cite{backdoor} and prompt injection attacks \cite{injection}. This involves creating stricter, verifiable rules for its API interactions to defend against threats like prompt injection and ensure its operational integrity in mission-critical environments.

\section{Conclusion}\label{sec:conclusion}

\sysname is an agentic AI framework designed to provide fully automated, end-to-end security for 6G O-RAN environments. Our approach bridges the critical gap between passive threat detection and active, closed-loop response. The framework's design presents a novel blueprint for building trustworthy AI-driven network defenses by integrating RAG with trusted knowledge bases like MITRE FiGHT and incorporation of robust safety guardrails. Our evaluations confirm that MobiLLM can effectively identify threats from five types of cellular attacks reproduced in a testbed, orchestrate complex mitigation strategies, and significantly reduce response latency, demonstrating the feasibility and potential of autonomous security operations in next-generation networks.
\section*{Acknowledgements}\label{Acknowledgements} 

This research was supported by a Small Business Innovation Research (SBIR) Phase I award N6893625C0023 from the Naval Air Warfare Center (NAWC), the NSF convergence accelerator program under award ITE-2326882, and the Google Cloud Research Credits Program. 
\bibliographystyle{abbrv}
\bibliography{references,6gxsec_hotnets}

\end{document}